\documentclass[pra,aps,floats,floatfix,twocolumn,showpacs,longbibliography,superscriptaddress,notitlepage,preprintnumbers,showkeys,10pt]{revtex4-1}
\usepackage[USenglish]{babel}
\usepackage{amsmath,amssymb,amsfonts}
\usepackage{graphicx}
\usepackage{dcolumn}
\usepackage{bm}
\usepackage[colorlinks=true,linkcolor=blue,citecolor=red]{hyperref}
\usepackage[T1]{fontenc} 
\usepackage{lmodern}
\usepackage[usenames,dvipsnames]{color}
\newcommand{\rozmiarjeden}{0.485\textwidth}

\newcommand{\rozmiartrzy}{0.485\textwidth}

\begin{document}
\preprint{\emph{Submitted to:} Physical Review B}
\title{Extended Falicov-Kimball model: exact solution for the ground state}
\author{Romuald Lema\'nski}
\email[e-mail: ]{r.lemanski@int.pan.wroc.pl}
\affiliation{Institute of Low Temperature and Structure Research, Polish Academy of Sciences, ul. Ok\'olna 2, PL-50422 Wroc\l{}aw, Poland}
\author{Konrad Jerzy Kapcia}
\email[Corresponding author; e-mail: ]{konrad.kapcia@ifpan.edu.pl}
\affiliation{Institute of Physics, Polish Academy of Sciences, Aleja Lotnik\'ow 32/46, PL-02668 Warsaw, Poland}
\affiliation{Institute of Nuclear Physics, Polish Academy of Sciences, ul. E. Radzikowskiego 152, PL-31-342 Krak\'{o}w, Poland}
\author{\fbox{Stanis\l{}aw Robaszkiewicz}}
\thanks{Deceased, 7 June 2017}
\affiliation{Faculty of Physics, Adam Mickiewicz University in Pozna\'n, ul. Umultowska 85, PL-61614 Pozna\'n, Poland}
\date{November 3, 2017}
\begin{abstract}
  The extended Falicov-Kimball model is analyzed exactly in the ground state at half filling in the limit of large dimensions.
  In the model the on-site and the intersite density-density interactions between all particles are included.
  We determined the model's phase diagram and found a discontinuous transition between two different charge-ordered phases.
  Our analytical calculations show that the ground state of the system is insulating for any nonzero values of the interaction couplings.
  We also show that the dynamical mean-field theory and the static broken-symmetry Hartree-Fock mean-field approximation give the same results for the model at zero temperature.
  In addition, we prove using analytical expressions that at infinitesimally small, but finite, temperatures the system can be metallic.
\end{abstract}


\maketitle


\section{Introduction}
\label{sec:intro}

In condensed-matter physics, correlations between electrons give rise to many intriguing phenomena ranging from simple band renormalization to complex phase diagrams with charge, spin, or orbital ordering as well as superconductivity \cite{MicnasRMP1990,ImadaRMP1998,YoshimiPRL2012,FrandsenNatCom2014,CominScience2015,NetoScience2015,CaiNatPhys2016,HsuNatCom2016,PelcNatCom2016,ParkPRL2017,NovelloPRL2017}.
The most popular model for the description of correlation effects in lattice systems is the Hubbard model (HM) \cite{HubbardPRSL1963}, which captures essential physics related to the competition between electron
localization (driven by the on-site $U$ interaction) and electron itinerancy \cite{GeorgesRMP1996,KotliarRmp2006}.
However, the tremendous effort of many researchers has resulted in exact results only for the dimensions
$D = 1$  \cite{NagaokaPRB1966,LiebPRL1968} and $D \rightarrow \infty$ \cite{GeorgesRMP1996}.
In the latter case, the calculations were made using the dynamic mean-field theory (DMFT), wherein for the HM it was usually necessary (except for some special cases; see, e.g,. outcomes obtained for the infinite-dimensional hyperperovskite lattice~\cite{NguenTranPRB2016}) to apply a numerical procedure to determine (approximately) the Green's functions.
It turns out, however, that the application of the DMFT formalism to the Falicov-Kimball model (FKM) results in an exact solution for all values of the interaction strength \cite{BrandtMielsch1989,BrandtMielsch1990,BrandtMielsch1991,BrandtUrbanek1992}.

The FKM is a simplified version of the HM, where only electrons with, e.g.,  spin down are itinerant \cite{FalicovPRL1969,RamirezPRB1970,ZlaticFreericksLemanskiCzycholl2001,FreericksRMP2003,FreericksBook2006,RibicPRB2016,CarvalhoPRB2013,CarvalhoPRB2014}.
Initially, it was introduced for a description of metal-insulator transitions in various transition-metal and rare-earth compounds, and since then, it has been intensively studied as a model of many other physical phenomena (for a review see Refs.~\cite{GruberMacris1996,Jedrzejewski2001,FreericksRMP2003,FreericksBook2006} and references therein).

The rigorous result for dimensions $D\geq2$ says that at low enough temperature the half-filled FKM possesses a long-range order; that is, the immobile electrons form the checkerboard pattern, the same as in the ground state \cite{KennedyPhysA1986,LiebPhysA1986,BrandtSchmidt1986,BrandtSchmidt1987}.
This result holds for arbitrary bipartite (alternate) lattices and for all values of the interaction strength $U$.
For $D\rightarrow\infty$  it was also shown independently  using the DMFT that the ordered charge-density-wave phase can occur at finite temperature \cite{FreericksPRB2000,ChenPRB2003,HassanPRB2007,MatveevPRB2008,Lemanski2014}.
It is worthwhile to notice here that the Monte Carlo simulations performed for $D=2$ systems also give results similar to those obtained within the DMFT \cite{MaskaPRB2006,ZondaSSC2009,AntipovPRL2016}.

The basic versions of both the HM and FKM include only local (on-site) interaction $ U $, but in real systems the Coulomb repulsion $ V $ between electrons located in neighboring lattice sites can be quite significant \cite{HubbardPRSL1963}, as it can lead to a change in even the nature of the metal-insulator phase transition \cite{HubbardPRSLA1964,MottBook1974}.
Moreover, the direct competition of local and nonlocal interactions captures both the effects of strong correlations and the tendency of the system to form inhomogeneous charge distributions.
This is why the effects of intersite Coulomb interactions  have been intensively studied in the extended HM (EHM) (e.g. Refs. \cite{AmaricciPRB2010,HuangPRB2014,KapciaPRB2017,TerletskaPRB2017,AyralPRB2017} and references therein).
Some studies on the subject have also been  performed for the extended version of the FKM (EFKM);
e.g., Refs. \cite{GajekJMMM2004,Farkasovsky2008,HamadaJPhysSocJap2017}.

However, very few rigorous results have been reported so far on the effects of intersite interactions studied within the EFKM. In fact, only Refs. \cite{DongenPRL1990,DongenPRB1992} reported exact results for this model when $ D \rightarrow \infty $, but they referred  to only the limiting cases of $U\rightarrow 0$ and $U\rightarrow \infty$.
In addition, the method used in Refs. \cite{DongenPRL1990,DongenPRB1992} of summing up over Matsubara frequencies did not allow  approaching the lowest temperatures, including $T=0$.

Here our goal is to present the exact solution of the EFKM at $T = 0$  in the whole range of interaction couplings. The solution allows us to resolve any doubts concerning, among other things, the nature of the ordered phase, the width of the energy gap for a given phase, and the kind of phase transition between different phases (continuous or discontinuous)
when $U$ or $V$ changes.
We actually use formulas for the Green's functions derived in Refs. \cite{DongenPRL1990,DongenPRB1992}, but instead of summing over Matsubara's frequencies, we determine the total energy of the system and other quantities directly from the integration of the density of states multiplied by the corresponding functions.
The results obtained allowed us to verify, refine, and extend the outcomes obtained in Refs. \cite{DongenPRL1990,DongenPRB1992}.
In particular, we found exactly that the system is an insulator in the ground state for any value of $ V> 0 $ and if $V=0$ for any $ U \neq 0 $ (for both $ U> 0 $ and $ U <0 $), including the special case $ V / U = 1/2 $ for which some ambiguities were signalized \cite{DongenPRB1992}.

In this paper we present the exact solution for the ground state of the half-filled EFKM
for any value of the on-site $U$ and the nearest-neighbor $V$ interaction strengths.
We obtain these results for the Bethe lattice in the limit of high dimensions, which allows us to obtain  exact analytical formulas for the total energy of the system $E (U, V)$ and for the difference in density of itinerant electrons on the neighboring lattice sites $d_1 (U, V)$.
In addition, we examine this model with the Hartree-Fock mean-field approximation (HFA), and we show that the analytical formulas for $ E (U, V) $ and $ d_1 (U, V) $ obtained with these two methods coincide with each other.
This result shows that at $T=0$  the DMFT and the standard HFA are equivalent,
what of course, is not true for finite temperatures.

The rest of the paper is organized as follows.
In Sec.~\ref{sec:modelandmethods} the model considered is presented (Sec.~\ref{sec:model}), and the equations are determined with two different methods (DMFT in Sec.~\ref{sec:DMFT} and HFA in Sec.~\ref{sec:HFA}).
Section~\ref{sec:results} is devoted to a discussion of analytical and numerical solutions.
Finally, in Sec.~\ref{sec:conclusions} we summarize  the results of this work and provide some future perspectives.

\section{Model and methods}
\label{sec:modelandmethods}

\subsection{Model}
\label{sec:model}

The conventional simplified HM, also known as the spin-less FKM \cite{FalicovPRL1969,RamirezPRB1970,FreericksRMP2003}, describes itinerant electrons and  localized ions, where only local (on-site) interactions between (itinerant) electrons and ions (localized electrons) occur.
On the other hand, the EFKM also contains  the Coulomb interactions $V$ between all the particles on adjacent sites of the crystal lattice \cite{DongenPRB1992}.
In the present work, we use the same Hamiltonian as van Dongen~\cite{DongenPRB1992}.
It is composed of the following four terms:
\begin{eqnarray}
\label{eq:ham}
H=H_t+H_U+H_V+H_{\mu},
\end{eqnarray}
where
\begin{eqnarray*}
H_t&=&\frac{ t}{\sqrt{Z}}\sum\limits_{\left\langle i,j \right\rangle}(c^+_{i\downarrow} c_{j\downarrow}+c^+_{j\downarrow} c_{i\downarrow}),
\quad H_U=U\sum\limits_{i}n_{i\uparrow}n_{i\downarrow},  \\
H_V&=&\frac{2V}{Z}\sum\limits_{\left\langle i,j \right\rangle,\sigma,\sigma '}n_{i\sigma}n_{j\sigma '},
\quad H_{\mu}=-\sum\limits_{i,\sigma}\mu_{\sigma}n_{i\sigma},
\end{eqnarray*}
with $Z$ being the coordination number.
$n_{i\sigma}$ is the occupation number, and $c^+_{i\sigma}$ ($c_{i\sigma}$) denotes the creation (annihilation) operator of an electron with spin $\sigma=\uparrow,\downarrow$.
Electrons with spin $\sigma=\uparrow$ are localized.
The prefactors in $H_t$ and $H_V$ have been chosen such that they yield a finite and non vanishing contribution to the free energy per site in the limit $Z\rightarrow\infty$.
$\left\langle i,j \right\rangle$ denotes the sum over nearest-neighbor pairs.
At half filling, i.e., $n=1$ ($n=  \tfrac{1}{L} \sum_{i,\sigma} \left\langle n_{i\sigma} \right\rangle $, where $L$ ---is the number of lattice sites), the chemical potential $\mu$ for both types of electrons is given by $\mu \equiv \mu_\sigma = \tfrac{1}{2}U+2V$ \cite{DongenPRB1992}.

\subsection{Dynamical mean-field theory}
\label{sec:DMFT}

The dynamical mean field theory is an exact approach for interacting fermion systems, including the EFKM, in the limit of high dimensions \cite{FreericksRMP2003}.
In this limit the nonlocal interaction term $V$ is treated at the Hartree level because the exchange (Fock)  and the correlation energy due to the intersite term are negligible~\cite{MullerHartmannZPB1989,MetznerVollhardtPRL1989,MetznerZPhysB1989,KapciaAPPA2016}.

The basic quantity calculated within the DMFT is the retarded Green's function $G(U,V,d_1;\varepsilon)$ defined for the complex $z$ with $\textrm{Im}(z)>0$.
Since we are dealing with a system composed of two sublattices, we need to determine two Green's functions, $G^+$ and $G^-$, separately for the $+$ and $-$ sublattices.
Here we use the Green's functions derived by van Dongen for the EFKM on the Bethe lattice in the limit of large dimensions \cite{DongenPRB1992}.
The formulas have the following form (for $t=1$):
 \begin{eqnarray}
       G^+(z)=\frac{z+v+\frac{1}{2}Ud -G^-(z)}{[z+v+\tfrac{1}{2}U -G^-(z)]
       [z+v-\tfrac{1}{2}U -G^-(z)]} \nonumber \\
\label{eq2B2:greenfuntions}	\\
       G^-(z)=\frac{z-v-\frac{1}{2}Ud -G^+(z)}{[z-v+\tfrac{1}{2}U -G^+(z)]
       [z-v-\tfrac{1}{2}U -G^+(z)]}, \nonumber
  \end{eqnarray}
where $v=V(d+d_1)$; $d$ stands for the order parameter, which is equal to the difference of mean values of the ion occupations on the $+$ and $-$ sublattices ($d=\langle n_i^+\rangle -\langle n_i^-\rangle$),
whereas $d_1$ is the difference of mean values of electron occupations of the sublattices  ($d_1=\langle n_e^+\rangle -\langle n_e^-\rangle$).
In fact, $d$ and $d_1$ are not independent quantities, because for a given $d$ the value of $d_1$ can be determined unambiguously (excluding the case of the coexistence of two phases, which is discussed further).
However, $d$ needs to be found from the condition for a minimum of the free energy.

One can define a staggered magnetization of the system as $m_Q = \tfrac{1}{2}(d-d_1)$ and a charge polarization $\Delta_Q = \tfrac{1}{2}(d+d_1)$.
Notice that due to the equivalence of the two sublattices the state with order parameters $\Delta_Q$ and $m_Q$ ($d$ and $d_1$) is equivalent to the state with parameters of opposite signs (i.e., in which they are equal to $-\Delta_Q$ and $-m_Q$ (-$d$ and $-d_1$), respectively).

\begin{figure}
	\includegraphics[width=\rozmiarjeden]{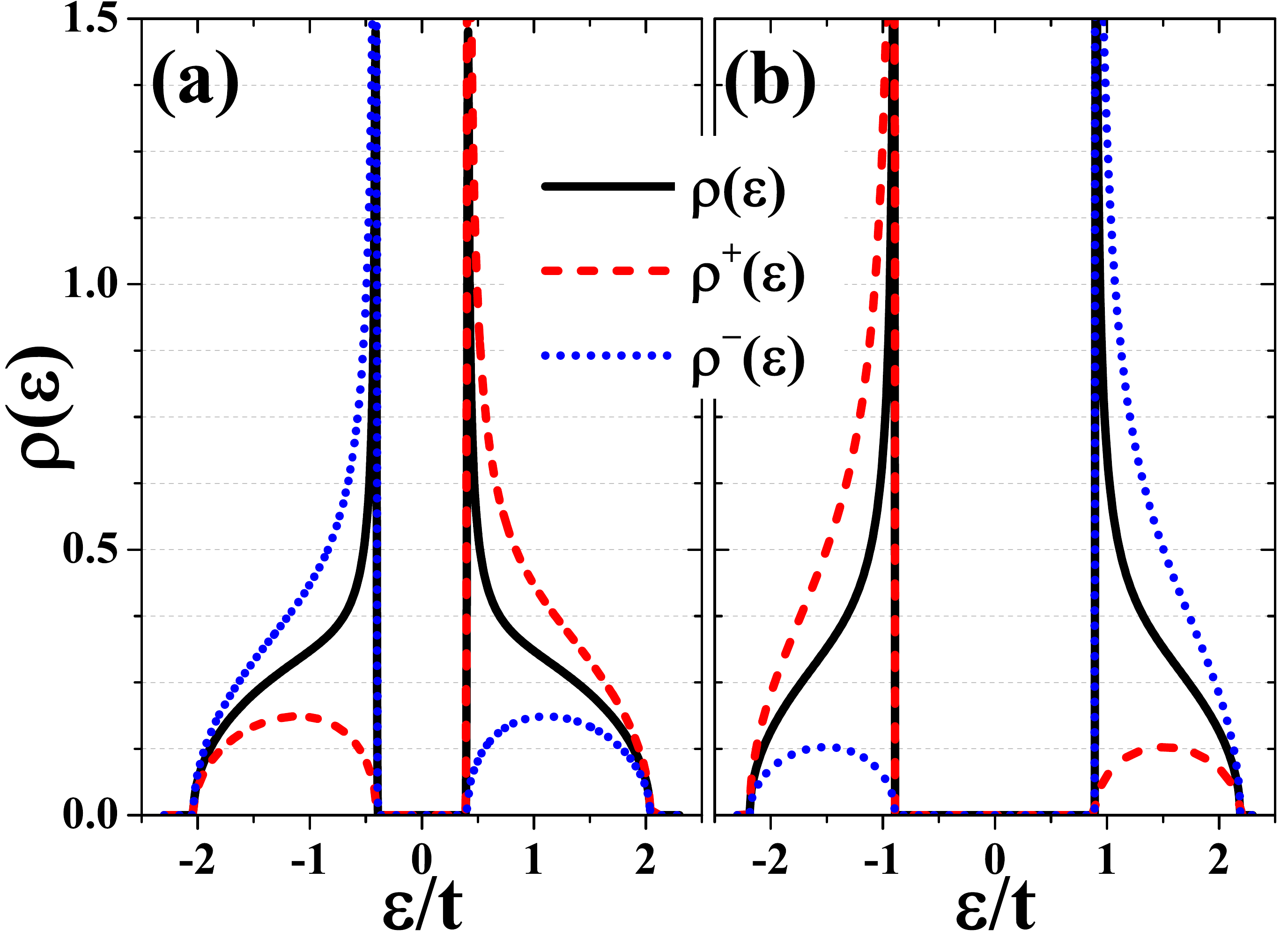}
	\caption{%
		The itinerant electron densities of states $\rho+$ (dashed line) and $\rho^-$ (dotted line) in each sublattice and total density of states $\rho=(1/2)(\rho^++\rho^-)$ for
        $U/t=1.0$ and (a) $V/t=0.2$ ($d_1=-0.52$, $A=-0.808$) and (b) $V/t=0.8$ ($d_1=0.74$, $A=1.784$).
        The Fermi level is located at $\varepsilon=0$.
	}
	\label{fig:DOS}
\end{figure}

By solving the set of equations (\ref{eq2B2:greenfuntions}) we calculate the density of states  using the standard formulas
  \begin{equation}
       \rho^{\pm} (U,V,d,d_1;\varepsilon)=-\frac{1}{\pi} \textrm{Im}G^{\pm}(U,V,d,d_1;\varepsilon +i0).
       \label{eq2B3}
       \end{equation}
It appears that at $T=0$, i.e., for $d=1$, the expressions for $\rho^+$ and $\rho^-$ take the following exact analytical forms:
  \begin{equation}
	\rho^{\pm}(\varepsilon)=\frac{1}{\pi}\left|
        \textrm{Im}\left(\frac{\sqrt{[4\varepsilon^2-A^2][4\varepsilon^2-A^2-16]}}{4[2\varepsilon\pm A]}\right)
        \right|,
  \label{eq2B4}
  \end{equation}
where $A=2V(1+d_1)-U$,
but still $d_1$ (``hidden'' in the parameter $A$) needs to be determined self-consistently from the equation
\begin{equation}
d_1=\langle n_e^+\rangle -\langle n_e^-\rangle,
\label{eq2B5}
\end{equation}
where
\begin{equation}
\langle n_e^{\pm}\rangle=\int_{-\infty}^{\varepsilon_F}\rho^{\pm}(U,V,d_1,\varepsilon)d\varepsilon
\label{2B6}
\end{equation}
and, in our case, the Fermi level is located at $\varepsilon_F=0$.
Notice that $\rho^{\pm}(\varepsilon)$ is non zero if $\varepsilon \in \left(-\sqrt{A^2+16}/2,-|A|/2\right) \cup \left(|A|/2,\sqrt{A^2+16}/2 \right)$.
Accordingly, with Eq. (\ref{eq2B4}), the densities $\rho^\pm(\varepsilon)$ depend only on the parameter $A$.
For all $A\neq0$ their shapes are qualitatively the same, with the singularities at the edges of the gap.
Exemplary densities of states are presented in Fig.~\ref{fig:DOS} for both signs of $d_1$.

The total ground-state energy $E_{tot}$ is given by
\begin{equation}
E_{tot}=\int_{-\infty}^{0}\varepsilon\rho(U,V,d_1,\varepsilon)d\varepsilon+ \frac{1}{4}\left[U + V\left(3+ d_1^2\right)\right],
\label{eq2B7}
\end{equation}
where $\rho=(\rho^{+}+\rho^{-})/2$.
The last term in (\ref{eq2B7}) is the sum of two constants, $U/4+V$, minus the expression $V(1-d_1^2)/4$, representing the interaction energy between moving electrons occupying adjacent lattice sites.
We need to subtract this expression because in the integral formula in the first part of (\ref{eq2B7}) the interaction is already computed twice.
And the sum $U/4+V$ is provided in order to normalize the total energy to the values consistent with those obtained both in Ref. \cite{DongenPRB1992} and by the HFA presented in the next section.

Since the functions $\rho^+$ and $\rho^-$ can be expressed only through a single parameter $A$, instead of three independent parameters $U$, $V$ and $d_1$, expressions  (\ref{eq2B5}) and  (\ref{eq2B7}) can be rewritten in the following form:
\begin{equation}
d_1  =  \frac{A}{2\pi}\int_{-\infty}^{0} d\varepsilon \frac{\sqrt{16+A^2-4\varepsilon^2}}{\sqrt{4 \epsilon^2 - A^2}},
\label{eq2B8}
\end{equation}
\begin{eqnarray}
	E_{tot} & = & \frac{1}{4}[U + V(3+ d_1^2)] \nonumber \\
	& - & \frac{1}{2\pi}\int_{-\infty}^{0} d \varepsilon \frac{\varepsilon^2 \sqrt{16+A^2-4\varepsilon^2}}{\sqrt{4 \varepsilon^2 - A^2}}
\label{eq2B9}
\end{eqnarray}
It appears that the integrals (\ref{eq2B8}) and (\ref{eq2B9}) can be expressed in analytical form using the elliptic integrals as follows.
\begin{equation}
 d_1 = \frac{A\sqrt{16 + A^2}}{4\pi}
\left[\mathbb{E}K\left(\tfrac{16}{16 + A^2}\right)-\mathbb{E}E\left(\tfrac{16}{16 + A^2}\right)\right]
\label{eq2B10}
\end{equation}
\begin{eqnarray}
E_{tot} & = & \frac{1}{4}[U + V(3+ d_1^2)]
 -  \frac{|A|}{48\pi} \left|\textrm{Im}\left\{32 \mathbb{E}K \left(\tfrac{16 + A^2}{A^2}\right) \right.\right. \nonumber \\
& - & \left.\left. (16 - A^2)\mathbb{E}E\left(\tfrac{16 + A^2}{A^2}\right)\right\}\right|,
\label{eq2B11}
\end{eqnarray}
where we used the following abbreviations: $\mathbb{E}K(x)=\textrm{EllipticK}(x)$ denotes the complete elliptic integral of the first kind and
$ \mathbb{E}E(x)=\textrm{EllipticE}(x)$ is the complete elliptic integral of the second kind.

Since at the Fermi level $\varepsilon_F=0$ the densities of states $\rho^{+ }$ and $\rho^{-}$ expressed by (\ref{eq2B4}) are both equal to zero,
the system is in the insulating state at $T=0$ for any non zero $U$ or $V$.
However, that is not always the case for $T>0$, when the order parameter $d<1$ (see, e.g., Ref.~\cite{Lemanski2014}).

\subsection{Hartree-Fock approach}
\label{sec:HFA}

The standard broken-symmetry HFA employing Bogoliubov transformation (including the long-range commensurate magnetic and charge orders, in agreement with the approach presented in Sec.~\ref{sec:DMFT}) also gives, at $T=0$, that $|d|=1$.
In this approach one gets the dispersion relation for itinerant quasiparticles $E_{quasi}=\pm \tfrac{1}{2}\sqrt{4\varepsilon + A^2}$, and energies for localized quasiparticles are $E_{loc}=\pm\tfrac{1}{2}[2V(1+d_1)-Ud_1]$, where $A= \left[ 2V(1+d_1) -U \right] $ as defined previously.
Notice that the spectrum of the itinerant quasiparticles has a gap $\Delta(\varepsilon_F)=|A|$, with the Fermi level located at the center of the gap ($\varepsilon_F=0$).

The equation for the parameter $d_1$ (we have assumed that $d>0$ and $t=1$) has the form
\begin{eqnarray}
\label{eq:HFAd1}
	d_1 & = & \frac{A}{\pi}\int_{-\infty}^{0} \frac{d \varepsilon \sqrt{4-\varepsilon^2} }{\sqrt{4 \varepsilon^2 + A^2}}.
\end{eqnarray}
The energy of the system per site is derived as
\begin{eqnarray}
\label{eq:HFAEtot}
	E_{tot} & = & \frac{1}{4}(U + V(3+ d_1^2)) \nonumber \\
	& - & \frac{1}{4\pi} \int_{-\infty}^{0} d\varepsilon \sqrt{4-\varepsilon^2}\sqrt{4\varepsilon^2 + A^2}.
\end{eqnarray}

On the other hand, the expressions for $d_1$ and $E_{tot}$ derived within the HFA
and given in (\ref{eq:HFAd1}) and (\ref{eq:HFAEtot}) are as follows:
\begin{equation}
\label{eq:HFAd1dwa}
  d_1  = \frac{A}{4\pi|A|} \left[(16+A^2) \mathbb{E}K \left( y \right) \right.
 \left. -A^2 \mathbb{E}E \left( y \right)\right]
\end{equation}
\begin{eqnarray}
\label{eq:HFAEtotdwa}
E_{tot} & = & \frac{1}{4}\left[U + V(3+ d_1^2)\right]    \\
& - & \frac{|A|}{48\pi} \left| (A^2+16)\mathbb{E}K\left( y \right) -(A^2-16) \mathbb{E}E\left( y \right) \right|, \nonumber
\end{eqnarray}
where $y=-16/A^2$ ($y<0$).

In the case of the Bethe lattice the noninteracting density of states is a semielliptic one, and it is expressed as $\rho_0(\varepsilon)=1/(2\pi t)\sqrt{4t^2-\varepsilon^2}$.
This expression with $t=1$ was used to derive Eqs.  (\ref{eq:HFAd1}) and (\ref{eq:HFAEtot}).

Before we  analyze the ground state of model (\ref{eq:ham}) we would like to comment the equations for $d_1$ and $E_{tot}$ derived within the DMFT and the HFA.
Although  Eqs. (\ref{eq2B10})  and (\ref{eq:HFAd1dwa}) for $d_1$ as well as Eqs. (\ref{eq2B11})  and (\ref{eq:HFAEtotdwa}) for $E_{tot}$ are not in the same analytical form, we have checked numerically that the functions $d_1(A)$ and $E_{tot}(A)$ obtained within both methods are the same (with a relative accuracy error of the order of $10^{-50}$).
Thus, the solutions found at $T=0$ for $d_1$ and $E_{tot}$ are the same for both approaches (for any $U\neq0$ and $V\geq0$).

This result is quite surprising because it is a well-known fact that the static mean-field theory is usually not an adequate tool for describing correlated electron systems.
According to the results of Ref.~\cite{MullerHartmannZPB1989},  both approaches used in this work (the DMFT and the HFA) for $U=0$ obviously give the same results (at any $T\geq0$) because the mean-field decoupling of the intersite terms is an exact one in the limit of high dimension (notice that model (\ref{eq:ham}) for $U=0$ is equivalent to the spinless fermion model; see, e.g., Refs.~\cite{UhrigPRL1993,UhrigAP1995}).
In the general case of $U\neq0$ the on-site correlations are often not properly captured by the HFA, and the  DMFT approach should be used for proper description of the correlated system, even at $T=0$, particularly in the case of the HM and the EHM~\cite{GeorgesRMP1996,KapciaAPPA2016}.
As we show in this work, the essential physics of the ground state of the half-filled EFKM is captured by the HFA, which gives a valid description for the long-range order of the itinerant as well as immobile particles.
For the half-filled FKM ($V=0$) the equivalence between the HFA and the DMFT was shown, e.g., in Ref.~\cite{NguyenTranPRB2013}.

\section{Ground-state results}
\label{sec:results}

\subsection{Ground-state phase diagram}
\label{sec:phasediagram}

\subsubsection{Charge-ordered and magnetic solution}

\begin{figure}
	\includegraphics[width=\rozmiarjeden]{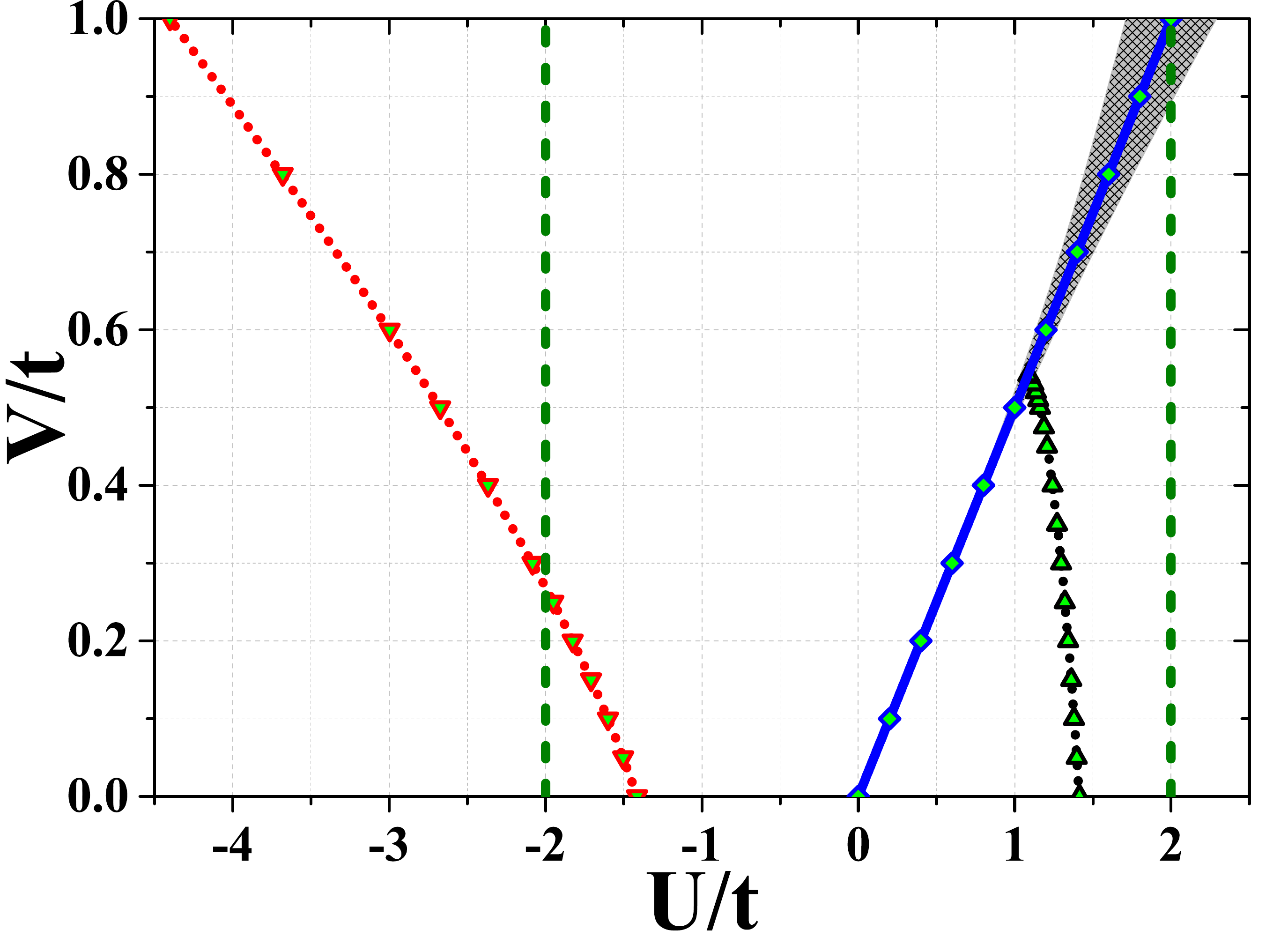}
	\caption{%
		Ground-state phase diagram of the model at half filling.
		Solid and dotted lines denote lines of quantum-critical points and \emph{quasi}-quantum-critical points, respectively.
        The quantum-critical points are associated with first-order transition between insulating phases with $dd_1>0$ (for $2V>U$) and $dd_1<0$ (for $2V<U$).
		The shaded area denotes a coexistence region, where these two phases can coexist.
		The dashed lines at $|U|/t=2$ denote the Mott metal-insulation transition with increasing $|U|/t$ in an absence of charge order (nonordered solution of the model).
	}
	\label{fig:GSphasediagram}
\end{figure}

The ground state of the model (\ref{eq:ham}) at half filling is insulating, and it is ordered with $|d|=1$ for any $U/t$ and $V/t$ (but, of course, the singular point $U=V=0$, when $d=d_1=0$ and the system is metallic).
The phase diagram of the model at $T=0$ is shown in Fig.~\ref{fig:GSphasediagram}.
For $2V>U$ one gets $dd_1> 0$, and the charge order dominates (i.e., $|\Delta_Q|>|m_Q|$); for $2V<U$ one gets $dd_1 < 0$, and antiferromagnetic order dominates ($|\Delta_Q|<|m_Q|$).
At $2V=U$ (for any $V>0$) there is a discontinuous transition between these two phases (the line $2V=U$ is a line of quantum-critical points; see Sec.~\ref{sec:firstordertransition}).
In the neighborhood of the transition line the coexistence region is also shown.
For small $V$ the region is narrower than the thickness of the line, but it is always finite for any $V>0$.
We also determine a location of the so-called quasicritical points, which are discussed later in Sec.~\ref{sec:quasicritical}.

\begin{figure}
	\includegraphics[width=\rozmiarjeden]{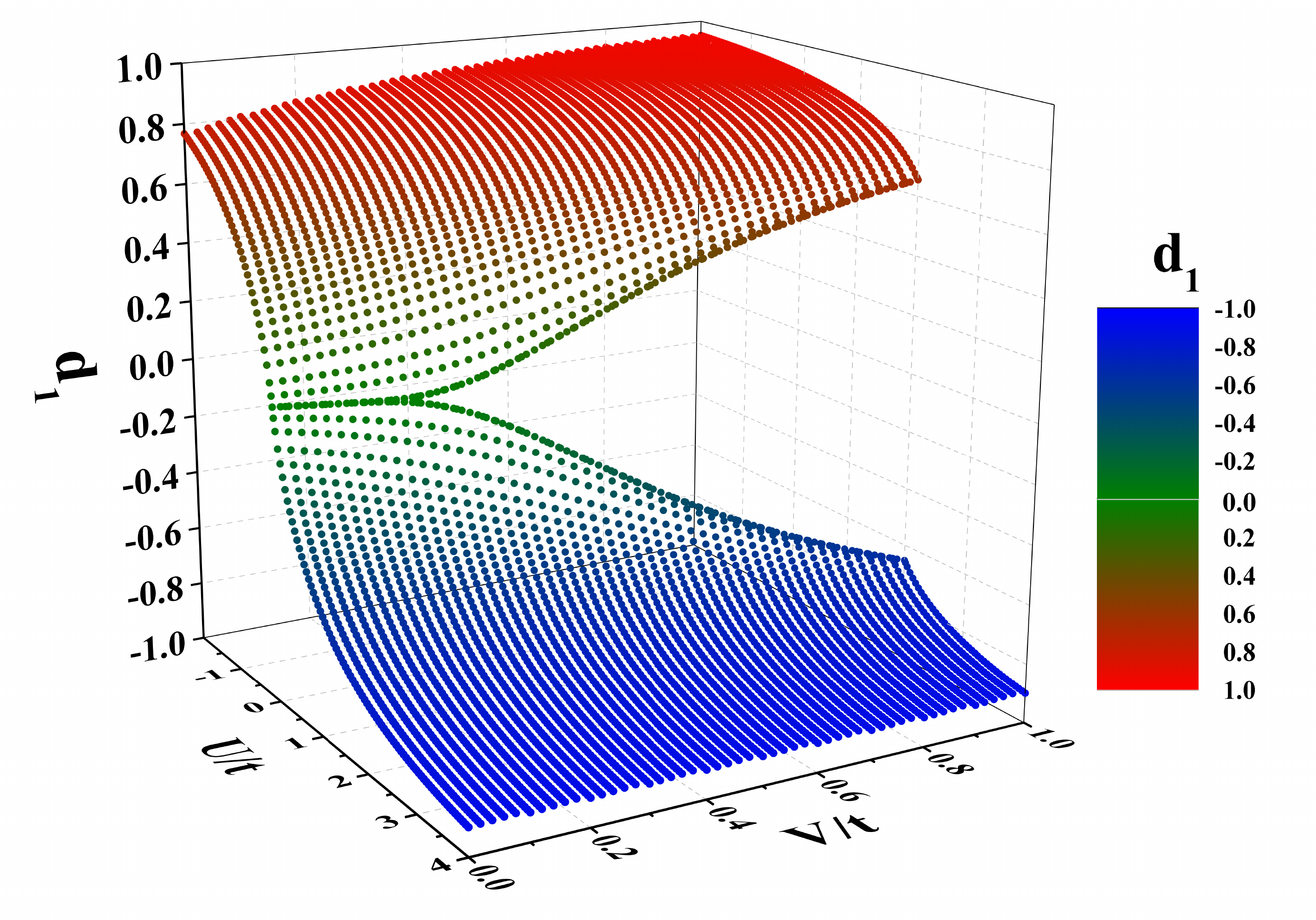}
	\caption{%
		The overall behavior of the parameter $d_1$ as a function of $U/t$ and $V/t$ in the ground state ($d=1$) and in the stable phases.
		The discontinuity at $U=2V$ is clearly visible.
		The color scale is included for better readability.
	}
	\label{fig:d1allUV}
\end{figure}

Figure~\ref{fig:d1allUV} presents the overall behavior of the parameter $d_1$ in the stable phases.
Notice that $|d_1|\rightarrow 1$ if $|U|\rightarrow\pm\infty$ or $V\rightarrow+\infty$ due to the fact that in these limits the model is equivalent to the extended Hubbard model in the zero-bandwidth limit (see, e.g., Refs.~\cite{MicnasPRB1984,KapciaPhysA2016,KapciePRE2017} and references therein).

\begin{figure}
	\includegraphics[width=\rozmiartrzy]{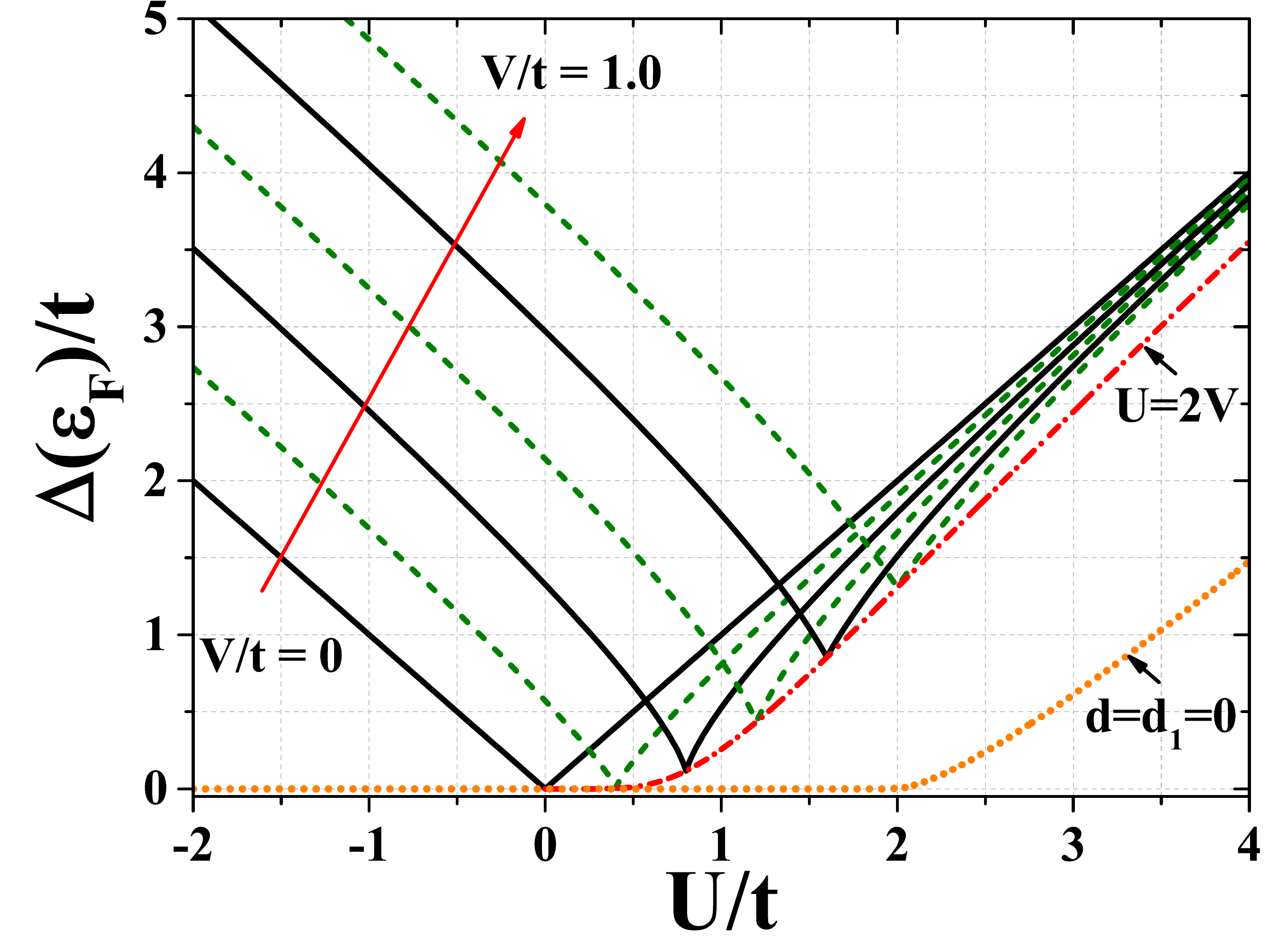}
	\caption{%
		The energy gap $\Delta(\varepsilon_F)$ at the Fermi level for itinerant electrons at $T=0$ (in the stable phases) as a function of $U$
		for a few values of $V$ ($V/t$ from a range from $0.0$ to $1.0$ with a step of $0.2$, from bottom to top; solid and dashed lines alternate).
		The dash-dotted line indicates a line of minimal $\Delta(\varepsilon_F)$ for fixed $V$ ($\Delta(\varepsilon_F)=2V|d_1|$ at $U=2V$).
		The dotted line presents the gap for the paramagnetic solution ($d=d_1=0$), which is independent of $V$.
	}
	\label{fig:energygap}
\end{figure}

It appears from Eq. (\ref{eq2B4}) that the energy gap $\Delta (\varepsilon _F)$ at the Fermi level (at $T=0$) is equal to the absolute value of the parameter $A$, so we have
\begin{equation}
\Delta (\varepsilon _F)=|A|=|2V-U+2Vd_1|.
\end{equation}
Notice that this expression coincides with the result for the gap in the quasi-particle spectrum obtained within the HFA.
Figure~\ref{fig:energygap} shows the evolution of $\Delta (\varepsilon _F)$ as a function of $U$ for a few values of $V/t$.
The minimum value of $\Delta (\varepsilon _F)$ is attained at $2V=U$, and it is equal to $2V|d_1|$.
Since the value of $d_1$  at $U=2V$ is very small for $V<0.2$, then  $\Delta (\varepsilon _F)$ also has a very small  (but nonzero) value  at the boundary line, and it can be hardly noticed in Fig.~\ref{fig:energygap} (dash-dotted line).
In the limit of large interactions where $|d_1|\rightarrow1$ the gap $\Delta (\varepsilon _F)$ can be expressed as $\Delta (\varepsilon _F)=4V-U$ if $U<2V$ and $\Delta (\varepsilon _F)=U$ if $U>2V$.

The observation that both the DMFT and the HFA give the same results for ordered solutions at $T=0$ leads to the conclusion that the insulating behavior of the ground-state ordered phases originates from the long-range order.
Obviously, the order is due to the interactions, but the interactions are not the direct reason for the emergence of the gap at the Fermi level, as it is in the case of the nonordered phase (see below).

Note that such a situation does not always occur in the ordered states.
For example, in the extended Hubbard model the insulating behavior at quarter filling originates from both correlations and long-range order \cite{AmaricciPRB2010,KapciaPRB2017}.
Insulating properties of that phase result from the Mott localization in one of the sublattices.
The other charge-ordered insulating solution of the EHM, which is the half-filled one, has a mean-field nature with respect to the symmetry-breaking transition and the DMFT description of that charge ordered state closely resembles the (static) mean-field solution of the problem \cite{KapciaPRB2017}.

\subsubsection{Paramagnetic solution}

Within the DMFT formalism, we also derived a diagram containing only paramagnetic (non-ordered) solutions of model (\ref{eq:ham}), i.e., assuming that $d=0$ and $d_1=0$ (it can  also  be treated as a high-temperature solution of the model (\ref{eq:ham}) \cite{HubbardPRSLA1964,VelickyPR1968,DongenAP1997,Lemanski2014}).
Within this assumption interaction $V$ is irrelevant (it only shifts a value of the chemical potential for $n=1$ and changes the total energy), and the model (\ref{eq:ham}) can be reduced to the standard FKM.
In such a case the ground state of the model is metallic for $|U|/t<2$ and insulating for $|U|/t>2$ \cite{Lemanski2014,RamirezPRB1970}.
At $|U|/t=2$ the system exhibits a metal-insulator transition, which is independent of $V$.
The energy gap $\Delta(\varepsilon_F)$ in the paramagnetic insulating phase does not depend on $V$, and it is expressed as follows \cite{Lemanski2014}:
\begin{equation}
\frac{\Delta(\varepsilon_F)}{t}= \sqrt{10+\left(\frac{U}{t}\right)^2-2\frac{1+\sqrt{\left[1+2(U/t)^2\right]^3}}{(U/t)^2}}.
\end{equation}
For $|U|/t<2$ one gets $\Delta(\varepsilon_F)=0$.
The $U$ dependence of $\Delta(\varepsilon_F)$ in the paramagnetic solutions is also shown in Fig.~\ref{fig:energygap} (dash-dotted line).
In the limit $|U|\rightarrow+\infty$ one gets that $\Delta(\varepsilon_F)\rightarrow|U-2|$.
For any $U$ at $T=0$ the gap in the paramagnetic solutions is lower than that corresponding to the ordered phases.

The metal-insulator transition between paramagnetic phases is driven by strong correlations between electrons.
In such a case the HFA fails and does not catch the essential physics due to the fact that the itinerant electrons interact with the dynamical effective field originating from ions and other electrons.

We need to stress that at $T=0$ the paramagnetic phases have higher energies than the charge-ordered ones mentioned before and they are not \emph{true} ground-state phases (even for $V=0$) \cite{FreericksPRB2000,LemanskiPRL2002,ChenPRB2003,Lemanski2014}.
However, the analysis of the nonordered state provides an insightful picture of the Mott's physics.

\subsection{The discontinuous transition at \texorpdfstring{$U=2V$}{U=2V}}
\label{sec:firstordertransition}

\begin{figure}
	\includegraphics[width=\rozmiartrzy]{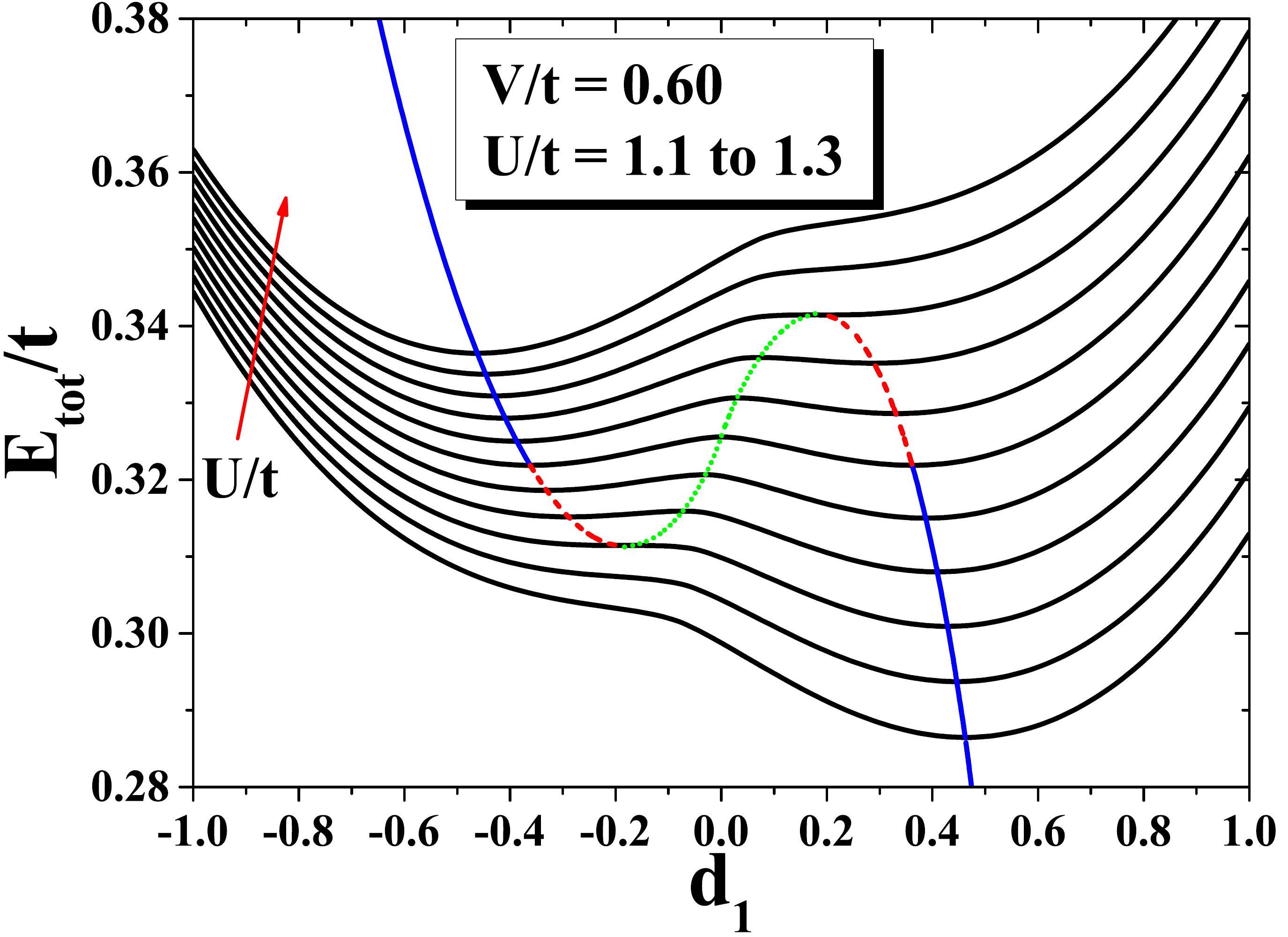}
	\caption{%
		The energy $E_{tot}$ as a function of $d_1$ for different values of $U/t$ (ranging from $1.1$ to $1.3$ with a step of $0.02$, from  bottom to  top).
		The dependence of $E_{tot}$ as a function of $d_1$ in stable, metastable, and unstable phases ($S$-shaped line; stable, metastable, and unstable solutions correspond to solid, dashed, and dotted lines, respectively) is also shown.
		The plot is derived for $d=1$.
	}
	\label{fig:V06propertiesEtotvsd1}
\end{figure}

\begin{figure*}
	\includegraphics[width=0.99\textwidth]{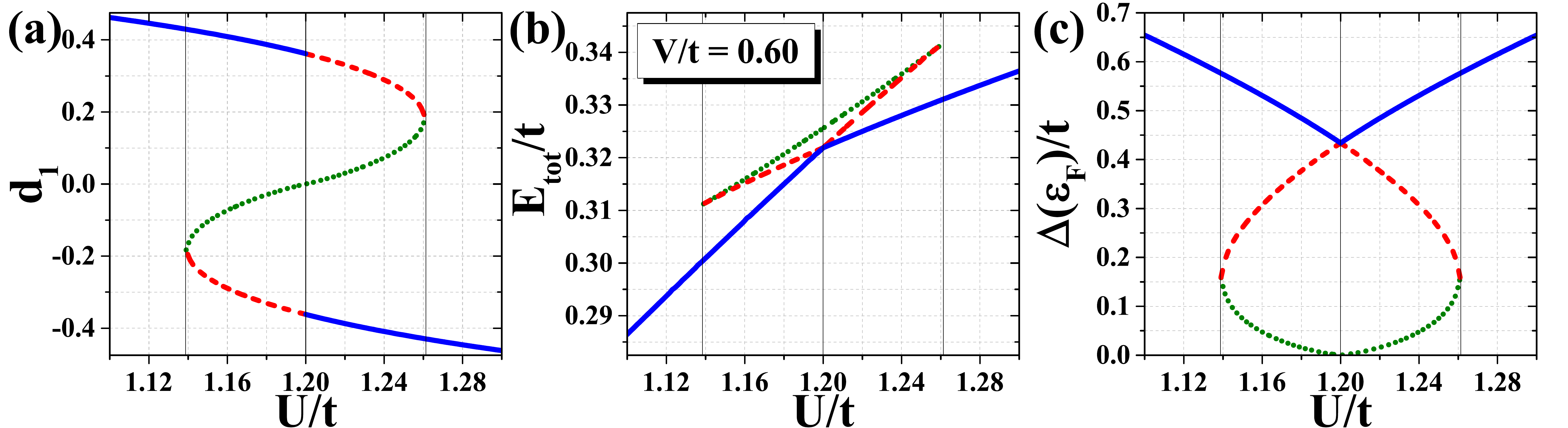}
	\caption{%
		The dependence of (a)  the parameter $d_1$, (b) energy $E_{tot}$, and (c) energy gap $\Delta(\varepsilon_F)$ at the Fermi level as a function of $U/t$ in stable, metastable, and unstable phases (denoted by solid, dashed, and dotted lines, respectively) for $V/t=0.60$ at the ground state.
		The plots are derived for $d=1$.
	}
	\label{fig:V06properties}
\end{figure*}

In this section we investigate in detail the behavior of the system in the  vicinity of the transition boundary at $U=2V$ and show that the transition is indeed discontinuous.

If for a given $V$ we insert  expression (\ref{eq2B4}) into (\ref{eq2B5}), it turns out that for $U=2V$ and in the vicinity of that value there exist three solutions for $d_1$.
Minimization of the ground state energy specified in (\ref{eq2B9}) shows that the intermediate $d_1$ solution  corresponds to the entirely unstable state, while one of two extreme solutions corresponds to the stable state, and the other corresponds to the metastable state.
Exactly at $U=2V $ the two minima of total energy have the same depth, and in the system a phase transition of the first kind occurs, which is accompanied by a jump of $d_1$ from a negative to a positive value when $U$ passes from $U>2V$ to $U<2V$.
Here, as an example,  we illustrate  in Fig.~\ref{fig:V06propertiesEtotvsd1} the situation for the case of  $V/t = 0.60$.
In Fig.~\ref{fig:V06propertiesEtotvsd1} the total ground-state energy $E_{tot}$ as a function of $d_1$ is shown for fixed values of $U/t$ (we consider only solutions with $d=1$).
$E_{tot}$ as a function of $d_1$, which is a solution of the set of equations (\ref{eq2B4})--(\ref{eq2B5}) (or equivalently Eq.~(\ref{eq2B8})), is also shown.
It is clearly visible that for $1.139<U/t<1.261$ the set has three solutions: two corresponding to the local minima of $E_{tot}(d_1)$ (stable and metastable solutions), and one associated with local maxima of  $E_{tot}(d_1)$ (unstable solution).
In this range of $U/t$ the two insulating phases (stable and metastable) can coexist (the coexistence region is denoted in Fig.~\ref{fig:GSphasediagram}).
At $U=2V=1.2$ we notice the discontinuous transition between the states with $d_1>0$ and $d_1<0$.
In Figs.~\ref{fig:V06properties}(a) and~\ref{fig:V06properties}(b) $d_1$ and $E_{tot}$ are presented as a function of $U/t$ in stable, metastable, and unstable phases.
In the proximity of the transition point at $U=2V$ the dependencies of $d_1$ and $E_{tot}$ exhibit the characteristic behavior expected in the neighborhood of the discontinuous transition.
Figure~\ref{fig:V06properties}(c) presents the gap $\Delta(\varepsilon_F)$  at the Fermi level in all solutions found.
Notice that $|d_1|$, and consequently, $\Delta(\varepsilon_F)$ in the unstable phase are smaller than those quantities in the stable and metastable phases.

\begin{figure}
	\includegraphics[width=\rozmiartrzy]{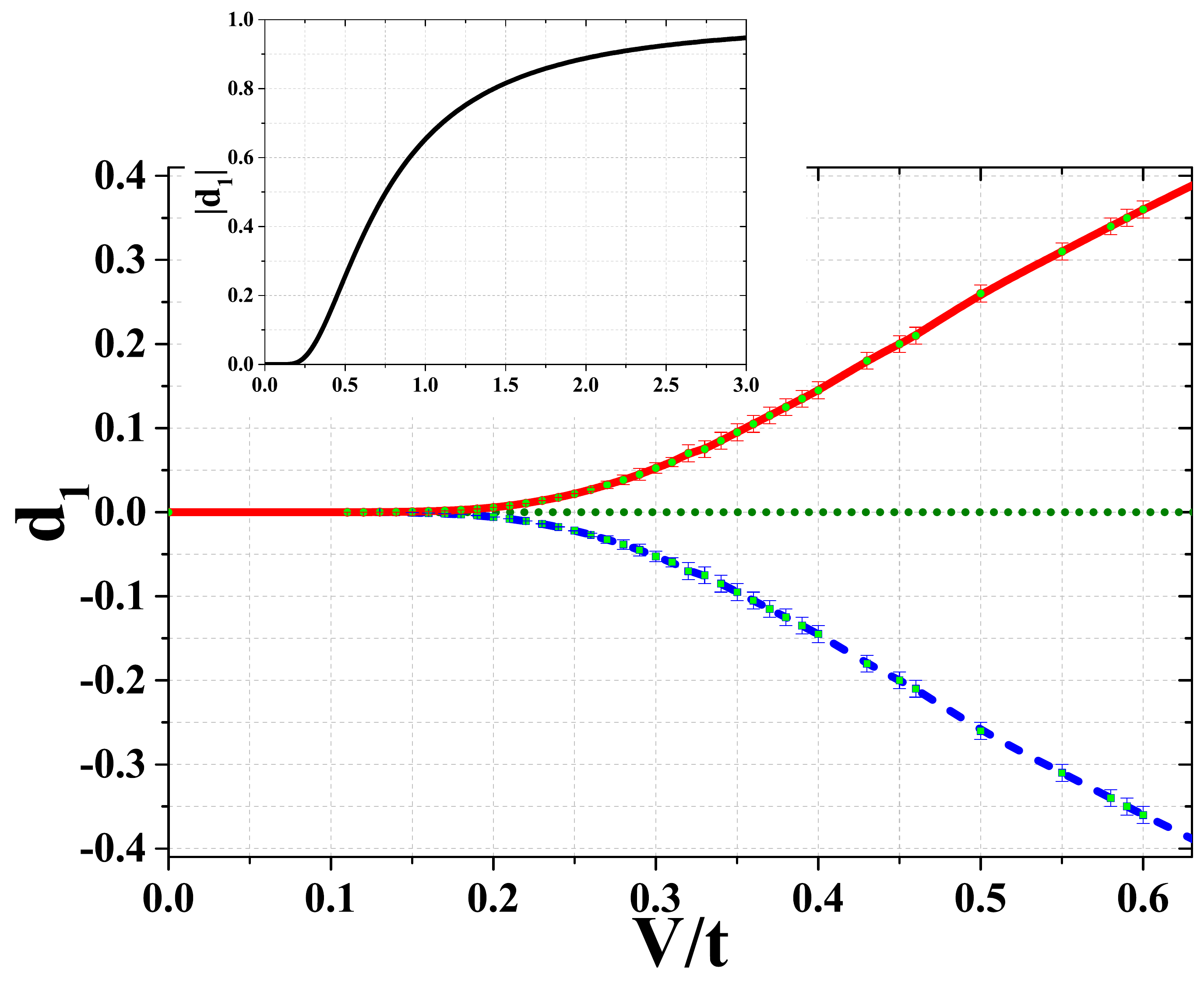}
	\caption{%
		The dependence of $d_1$ as a function of $V/t$ at the quantum-critical point line (for $U=2V$).
		The solid and dashed lines denote the solutions for both stable coexisting phases (with $d_1>0$ and $d_1<0$, respectively), whereas the dotted line denotes the unstable solution ($d_1=0$).
		The inset presents $|d_1|$ in both stable phases for larger $V/t$.
	}
	\label{fig:d1oncriticalline}
\end{figure}

When $V$ becomes small ($V<0.2$), the jump of $d_1$ at the phase-transition point $U=2V$ is extremely small
(see Figs.~\ref{fig:d1allUV} and~\ref{fig:d1oncriticalline}) and it is unclear whether it disappears.
However, the precise calculations showed that there is a finite jump of $d_1$ for any positive $V$, so the phase transition is always discontinuous for any $V>0$, and it changes into a continuous one only for $V=0$ \cite{Lemanski2014}.
With decreasing $V$ the coexistence region is also reduced gradually, but it is always finite.
At $U=V=0$ only one stable solution with $d_1=0$ exists, but this is a special singular point, as  the condition $d=1$ cannot be achieved, only $d=0$ (with $d_1=0$).

We should underline the fact that at $T=0$ the solution of the model with $d_1=0$ is found only for $U=2V$; however, as we already mentioned, it is entirely unstable for any $V>0$.
The solution was also found in Ref.~\cite{DongenPRB1992}, but its nature was not  investigated carefully.
The conclusion of that paper that the transition at $T=0$ and $U=2V$ is always continuous has been refuted convincingly by the analysis performed above.
Notice also that this (unstable for $V>0$, stable for $V=0$) solution is the only one, where there is no gap in the system at $T=0$, i.e., $\Delta(\varepsilon_F)=0$.

\subsection{\emph{Quasi-quantum-critical} points (\texorpdfstring{$T\rightarrow0^+$}{T->)+} limit)}
\label{sec:quasicritical}
Since the DMFT also allows us to study rigorously the system at finite temperatures, we find it useful to determine  the quasi-quantum-critical points in the phase diagram.
We use this name with respect to points where the density of states at the Fermi level $\rho(\varepsilon_F)$ is positive at arbitrarily low positive temperature, although at $T = 0$ we have $\rho(\varepsilon_F) = 0$.
In the case of $V = 0$ such points are found to be $U =\pm\sqrt2$ \cite{HassanPRB2007,Lemanski2014}.

It is quite fortunate that also for $V>0$  we can derive  from Eqs. (\ref{eq2B2:greenfuntions}) and (\ref{eq2B3}) the exact analytical formula for $\rho^{+}(\varepsilon_F =0)=\rho^{-}(\varepsilon_F =0)$ as a function of $U$, $V$, $d$, and $d_1$  (it is not possible to do so in the general case of any value of $\varepsilon$), which takes the following form:
\begin{equation}
\rho^{\pm}(U,V,d,d_1;\varepsilon_F)=\frac{1}{2\pi}\left|\frac{\textrm{Im}\left(\sqrt{w(U,V,d,d_1)}\right)}{U^2-4(d+d_1)^2V^2}\right|
\label{eq2B12}
\end{equation}
where $w(U,V,d,d_1)$ is the polynomial function:
\begin{eqnarray}
\label{eq2B13}
w(U,V,d,d_1) & = &  U^6-4U^4+4d^2U^2   \\
& + & 8dU(U^2-2)(d+d_1)V  \nonumber \\
& - & 8(U^4-2U^2-2)(d+d_1)^2V^2 \nonumber \\
& - & 32dU(d+d_1)^3V^3+16U^2(d+d_1)^4V^4  \nonumber
\end{eqnarray}
For the ground state, i.e., when $d=1$, the function $w(U,V,d=1,d_1)$ factorizes and takes the following form:
\begin{eqnarray}
\label{eq2B14}
w(U,V,d=1,d_1)= \qquad \qquad \qquad \qquad \qquad \qquad\\
\left[ U - 2 (1+d_1) V \right]^2 \left[U^2 + 2 U(1+d_1)V-2\right]^2 \nonumber
\end{eqnarray}
Consequently, $w(U,V,d=1,d_1)=0$ and $\textrm{Im}(w(U,V,d=1,d_1))=0$ for any $U$, $V$, and $d_1$.
Thus, we get $\rho^{\pm}(U,V,d=1,d_1;\varepsilon_F)=0$ (i.e. at $T=0$), as mentioned earlier.

Now we would like to find the values of $U_{qqcr}(V)$ for which the condition $w(U,V,d,d_1)<0$ is fulfilled for any $d<1$.
In other words, for a given $V$ we want to determine values of $U=U_{qqcr}$ for which
$\rho^{\pm}(U_{qqcr},V,d,d_1;\varepsilon_F)$ becomes positive for all $d<1$.

The simplest situation occurs for $V = 0$, because the polynomial $w(U,V,d,d_1)$ reduces to the form
$w(U,d)=U^2(U^4-4U^2+4d)$; hence, we get $U_{qqcr}(V=0)=\pm \sqrt{2}$.
This case was already considered in Ref. \cite{Lemanski2014}.

When $ V>0$, finding $U_ {qqcr}$ becomes more complicated because we have to determine first the value of the parameter $d_1$. But
from (\ref{eq2B14}) we see that $w(U,V,d=1,d_1)$ attains its minimum value equal to zero when $d_1=(2-U^2)/2UV-1$ or $d_1=U/2V-1$.
If we now insert the first of these two expressions into (\ref{eq2B5}) (the latter expression leads to an unphysical solution) and solve this self-consistent equation for a given $V$, then we get $U_ {qqcr}(V)$.
Indeed, if $d<1$, then for any $d_1$ the function $w(U_{qqcr},V,d,d_1)<w(U_{qqcr},V,d=1,d_1)$; hence, it becomes negative within a certain interval of $d_1$ around $d_1=(2-U^2)/2UV-1$, thus producing a positive value of the density of states at $\varepsilon_F$.

It appears that also for $V>0$ both positive and negative solutions for $U_ {qqcr}(V)$ exist.
In the ground state phase diagram in Fig.~\ref{fig:GSphasediagram} they are displayed by the dotted lines (black for $U>0$ and red for $U<0$).
For $V=0$ the positive and negative values of $U_{qqcr}$  are symmetrically distributed around $U=0$ and equal $\sqrt2$ and $-\sqrt2$, respectively, but  for $V>0$ their positions around $U=0$ becomes asymmetric.

An increase of $V$ moves $U_{qqcr}(V)$ towards smaller values  in the case of both $U>0$ and $U<0$.
However, for $U>0$ it diminishes smoothly down to the value $U_{qqcr}\approx 1.086$ for  $V\approx 0.543$,
where it meets the line $U=2V$.
So when $U$ is positive, the minimum value of $U_{qqcr}$ exists for the maximum value of $V\approx 0.543$, and for $V$ greater than $\approx 0.543$  there is no quasicritical point $U_{qqcr}$.
On the other hand, for $U<0$ there is apparently no minimum value of $U_{qqcr}$ and, consequently, no maximum value of $V$ above which there is no such quasicritical point (see Fig. 1).

It is obvious that the HFA cannot find the quasicritical point discussed in this section because it is a feature of the model at infinitesimally small, but finite temperatures and the dynamical effects which are captured by the DMFT are completely neglected in the HFA.

\section{Conclusions and final remarks}
\label{sec:conclusions}
In this work, we studied the ground-state behavior of the EFKM, which also includes  nonlocal interactions.
To investigate the system on the Bethe lattice the DMFT and the HFA (static mean field) were employed.

The main achievement of our work was finding the exact solution in the form of analytical formulas for the density of states, energy, and energy gap, depending on the coupling constants $U$, $V$ and $d_1$.
The parameter $d_1$, being the difference between the average densities of moving electrons at neighboring lattice sites, was determined in a self-consistent way from the integral (\ref{eq2B8}) or, equivalently, from Eq. (\ref{eq:HFAd1}).
Based on these equations, the ground-state phase diagram was determined, and it was proven that for any value of $V> 0$ the system is an insulator.
Moreover, the phase transition that occurrs when the coupling $U$ changes from $U <2V$ to $U> 2V$ is always discontinuous and is accompanied by the finite jump of $d_1$  from $d_1> 0$ to $d_1 <0$.
But if $U = 2V$, then the phases with $d_1<0$ and $d_1>0$, both of which are insulators, coexist, while the solution with $d_1=0$, which corresponds to the phase with zero energy gap, is  completely unstable.
We have therefore shown that the conclusion presented in \cite{DongenPRB1992}, that the phase transition is continuous when $U=2V$, is incorrect.

One of the most surprising conclusions drawn from all these studies is that
the solutions obtained with DMFT and the static HFA are equivalent at $T=0$.
It was proven in Ref.~\cite{MullerHartmannZPB1989} that in the limit of high dimensions the on-site $U$ interaction is the only one, which remains dynamical.
It was also shown rigorously that all other (intersite) interactions reduces to their Hartree approximations.
Thus, obviously, at any temperature (including $T=0$) the DMFT and the HFA give the same results for any $V>0$, but only for $U=0$.
The results of the present work show that  for the EFKM the HFA also gives the proper description of the ground-state properties of the system.
This result suggests that  the insulating behavior of the ground-state ordered phases originates from the long-range order occurring in the system, rather than directly from the interactions.
In contrast, the metal-insulator transition between paramagnetic phases is directly driven by interactions.

We also found that the ground state of the EFKM is not analytic continuation of the $T\rightarrow0^+$ states.
As we demonstrated analytically, the gapless (i.e., metallic) charge-ordered phase cannot occur at $T=0$, but it appears at infinitesimally  small temperatures at the quasi-quantum-critical point, i.e., for $U=U_{qqcr}(V)$, only if  $V\lessapprox 0.543$.
Indeed, $\Delta(\varepsilon_F)$ is not a continuous function of $d$ at $d = 1$ due to the appearance of the subgap bands for $d=1-0^+$.
We leave the details of the finite temperature studies of the model to subsequent publications.

Finally, let us comment that the EFKM analyzed in this work can be treated as a limiting case of the extended asymmetric HM, where both types of electrons can move but their hopping amplitudes are different from each other \cite{FathPRB1995}.
In the absence of intersite $V$ interactions in that model it was found that the disordered phase exhibits an  orbital-selective crossover at finite temperatures to the non-Fermi-liquid phase \cite{WinogradPRB2011,DaoPRA2012,WinogradPRB2012}.
We believe that an analytical result for the various generalizations of the EFKM similar to those presented in this paper can be found.

\begin{acknowledgments}
The authors express their sincere thanks to J. K. Freericks and M. M. Ma\'ska for useful discussions on some issues raised in this paper.
K.J.K. acknowledges the support from  the National Science Centre (NCN, Poland) under grant no. UMO-2017/24/C/ST3/00276.
\end{acknowledgments}

\bibliography{ESHMbibliography}

\end{document}